\documentclass[twocolumn,showpacs,preprintnumbers,prl,floatfix]{revtex4-1}
\usepackage{amssymb,amsmath,amsfonts}
\usepackage{graphicx,epsfig} 
\usepackage{bm} 
\usepackage[T1]{fontenc}
\usepackage{dcolumn}
\usepackage{varioref}
\usepackage{hyperref}
\usepackage{cleveref}
\usepackage{mathptmx}
\usepackage{relsize}
\usepackage{nicefrac}
\usepackage{subfigure}
\usepackage{xcolor}
\usepackage{booktabs}
\usepackage{comment}

\definecolor{Myorange}{cmyk}{0,0.42,1,0}

\begin{document}
\title{Hyperbolic embedding of brain networks detects regions disrupted by neurodegeneration in Alzheimer's disease} 
\author{Alice Longhena,$^{1}$ Martin Guillemaud,$^{1}$ Fabrizio De Vico Fallani,$^{1}$ Raffaella Migliaccio,$^{1,2}$ Mario Chavez$^{3}$}
\affiliation{%
$^1$~Paris Brain Institute, CNRS-UMR7225, Inserm-U1127, Sorbonne University-UM75, Inria-Paris. Piti\'e Salp\^etri\`ere University Hospital. Paris, France}%
\affiliation{%
$^2$~Institut de Neurologie, Piti\'e Salp\^etri\`ere University Hospital, AP-HP Paris, France}
\affiliation{%
$^3$~CNRS UMR7225, Piti\'e Salp\^etri\`ere University Hospital. Paris, France}
\date{\today}
\begin{abstract}
Graph-theoretical methods have proven valuable for investigating alterations in both anatomical and functional brain connectivity networks during Alzheimer’s disease (AD). Recent studies suggest that representing brain networks in a suitable geometric space can better capture their connectivity structure. This study introduces a novel approach to characterize brain connectivity changes using low-dimensional, informative representations of networks in a latent geometric space. Specifically, the networks are embedded in a polar representation of the hyperbolic plane, the hyperbolic disk. Here, we use a geometric score, entirely based on the computation of distances between nodes in the latent space, to measure the effect of a perturbation on the nodes. Precisely, the score is a local measure of distortion in the geometric neighborhood of a node following a perturbation. The method is applied to a brain network dataset of patients with AD and healthy participants, derived from diffusion-weighted (DWI) and functional (fMRI) magnetic resonance imaging scans. We show that, compared with standard graph measures, our method more accurately identifies the brain regions most affected by neurodegeneration. Notably, the abnormalities detected in memory-related and frontal areas are robust across multiple brain parcellation scales. Finally, our findings suggest that the geometric perturbation score could serve as a potential biomarker for characterizing the progression of the disease.

\end{abstract}

\maketitle %

\section{Introduction}
Network analysis of brain connectivity has advanced our understanding of the organizational mechanisms underlying different brain states. Interestingly, graph analysis of brain networks has also led to the development of biomarkers that quantify reorganization mechanisms in brain diseases~\cite{stam2014modern, manzoni2020network}.
In particular, neurodegenerative diseases like Alzheimer's disease (AD) have garnered significant attention from the network neuroscience community~\cite{tijms2013alzheimer, gomez2014network}.

Classic late-onset Alzheimer’s disease is a neurodegenerative disorder that causes progressive cognitive and functional impairments. It is histopathologically characterized by the appearance of amyloid-$\beta$ plaques~\cite{hampel2021amyloid} and tau-related neurofibrillary tangles~\cite{braak1991neuropathological, kowall1987axonal}, which lead to neuronal loss, brain atrophy, and disrupted normal inter-neuronal connectivity. This causes a gradual breakdown of cerebral tissue that disrupts both anatomical and functional communication between affected brain regions. The loss of long-range white matter bundles has been shown to impact anatomical brain connectivity in several areas, particularly the parietal and temporal lobes~\cite{agosta2011white, caso2015white}. At the whole connectivity level, these disruptions are evident from an increased characteristic path length and decreased communication efficiency (compared to healthy subjects) in brain networks estimated from diffusion-weighted imaging (DWI)~\cite{lo2010diffusion, pereira2018abnormal}. Similar connectivity alterations have also been documented in resting-state brain networks estimated from functional magnetic resonance imaging (fMRI)~\cite{wang2007altered, badhwar2017resting, dai2019disrupted}, as well as from magnetoencephalographic (MEG)~\cite{stam2009graph} and electroencephalographic (EEG) signals~\cite{de2009functional}.

According to the anatomopathological model proposed by Braak and Braak~\cite{braak1993staging}, the pathology typically spreads from the medial temporal areas, particularly from the entorhinal cortex, to the hippocampal formation. At later stages, it eventually progresses into the isocortex, from posterior to anterior regions.
Clinically, this spreading pattern corresponds to the initial appearance of episodic memory disturbances, followed by deficits in temporal-spatial orientation, language difficulties, praxic planning issues, and problems with object and face recognition. In later stages, disorders of the "frontal" sphere appear, leading to executive impairments and behavioral changes such as increased apathy, agitation, and inappropriate social behavior.
Consistent with histopathological and neuropathological models, network-based studies have detected connectivity changes in the temporal lobe, particularly in the hippocampus~\cite{wang2006changes, zhou2008abnormal}, during early AD stages, which reflect memory-related symptoms. Other studies report topological changes in the frontal regions~\cite{lo2010diffusion}, associated with behavioral deficits.
However, the abnormality of an AD biomarker (a measurable indicator of the state of the disease) does not always correlate with the severity of clinical symptoms~\cite{jack2010hypothetical}. Some biomarkers become abnormal before the associated symptoms appear, and these can be crucial for the diagnosis (diagnostic biomarkers) or progression tracking (progression biomarkers) of the disease.

In recent years, network models have shown great potential for embedding brain connectivity graphs into latent spaces where nodes are represented as low-dimensional vectors~\cite{cui2018survey, xu2021understanding} or probabilistic density functions~\cite{xu2020new, xu2021understanding}. Network embeddings are designed so that some properties of the original graph structure, usually measures of similarity and proximity between nodes, are maximally preserved in these spaces~\cite{goyal2018graph, xu2021understanding}. By representing networks as vectors or functions, a wide variety of machine learning algorithms can be applied to enable network visualization, link prediction, node classification, and node clustering~\cite{xu2021understanding}. Most current studies embed brain networks into Euclidean spaces to study their connectivity properties~\cite{rosenthal2018mapping, mach2023connectome}. However, such embeddings often require high-dimensional representations and fail to encompass some graph topological properties typical of complex real-world networks, such as hierarchical structure and high clustering~\cite{nickel2017poincare, whi2022hyperbolic}. In the case of tree-like structures, as the number of nodes grows exponentially with their distance from nodes of higher hierarchy, the dimensions required to correctly represent these nodes in Euclidean space increase considerably. Nevertheless, increasing dimensionality leads to greater computational complexity and high distortion of the representation~\cite{nickel2017poincare, chami2019hyperbolic}. In contrast, non-Euclidean (hyperbolic) mapping methods require reduced dimensionality to accurately embed graphs and better capture their scale-free and hierarchical structures~\cite{whi2022hyperbolic}.

Hyperbolic embedding has recently been found to encode, with minimal distortion and low computational cost, both local and global topological information of the original graphs~\cite{xu2020new}. The interest of the scientific community has thus shifted towards finding latent hyperbolic space models capable of explaining the emergence of complex network structures. Indeed, it has been demonstrated that heterogeneous degree distribution and high clustering emerge naturally from the metric definition of spaces with negative curvature~\cite{krioukov_hyperbolic_2010}.
When applied to anatomical brain networks from healthy subjects, hyperbolic embedding has revealed a multiscale connectivity structure between the anatomical brain regions~\cite{zheng2020geometric}. In a recent work, this geometric representation has shed new light on the characterization and localization of network perturbations produced by brain surgery in epilepsy~\cite{longhena2024detecting}. Similarly, a recent study on functional brain networks estimated from MEG signals has shown that hyperbolic embedding can be a useful tool to explore brain network perturbations associated with cognitive decline in Alzheimer's disease~\cite{baker2023hyperbolic}.

Here, we investigate the advantage of a geometric measure based on hyperbolic representations of brain (anatomical and functional) networks in identifying regions disrupted by neurodegeneration. Our approach involves the use of hyperbolic coalescent mapping to embed a graph in the hyperbolic disk~\cite{muscoloni_machine_2017}.
Given the embeddings, we can define new network metrics based on the relative distances between nodes in the latent geometric space. A geometric score has been recently proposed in~\cite{longhena2024detecting} to quantify the effect of a connectivity perturbation on a node, showing a good localization power and a small bias. In the present study, we test its ability to localize the most significant regional perturbations following connectivity disruption due to the progression of the disease. Finally, we assessed the role of the spatial scales (the number of nodes used to estimate the networks) in the identification of disrupted regions.

\begin{figure*}[!ht]
	\includegraphics[width=0.9\linewidth]{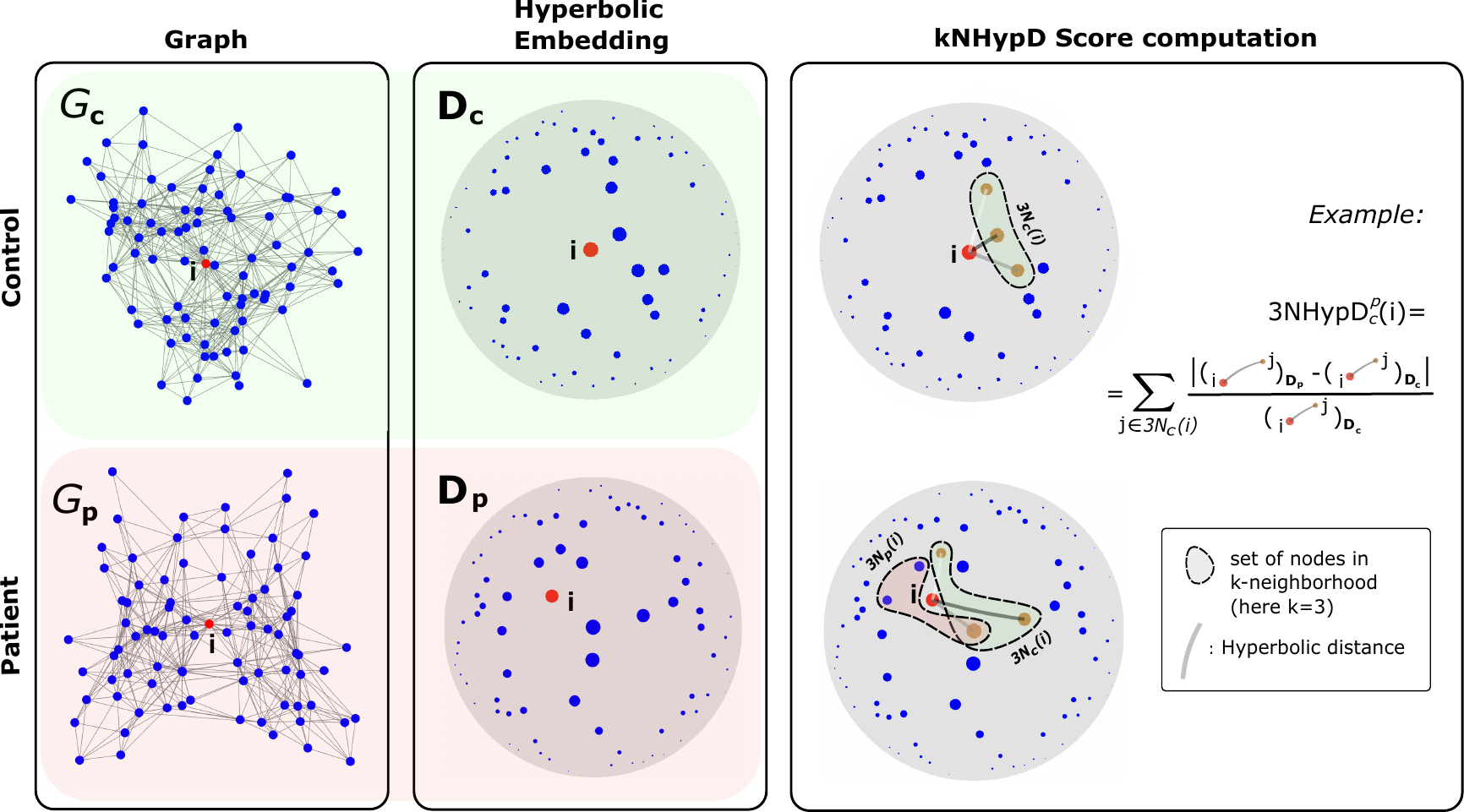}
	\caption{Illustration of the pipeline for the computation of $\mathrm{kNHypD}$ score for the comparison of two brain networks. From left to right: we take the filtered structural brain networks of a patient (graph $G_p$) and that of a control (graph $G_c$); we embed them in a hyperbolic latent space through coalescent embedding~\cite{muscoloni_machine_2017}, then rescale their radial coordinates to fit them in a unitary hyperbolic disk, the two embeddings are denoted as $\mathbf{D}_p$ and $\mathbf{D}_c$; as an example, we illustrate the computation of the local perturbation of the 3-neighborhood of node $i$ (marked by the red dot in the plots) with the score value $\mathrm{kNHypD}^{p}_c(i)$, with k$=3$. The set of nodes $3N(i)$ is indicated by the yellow points in the disks.}
	\label{fig:method}
\end{figure*}

\section{Methods}
\label{sec:methods}

\subsection{Data}
The dataset used for this study comprises 23 patients diagnosed with Alzheimer's disease and 25 healthy participants. For each of them, we have a functional brain network derived from functional magnetic resonance imaging (fMRI) at rest and a structural brain network derived from diffusion-weighted magnetic resonance imaging (DWI), representing the functional and structural connectivity at the time the scans were performed.
Demographic information and neurological characterization of the participants are reported in Table \ref{tab:data}. Additionally, $92\%$ of them reported being right-handed.
We display the results of the Mini Mental Status Exam (MMSE), a 30-point questionnaire to measure cognitive impairment, as well as the results of the free and cued selective reminding test~\cite{grober1987genuine,lemos2015free} for episodic memory. Specifically, this includes Total Recall (Free+Cued) (TR (F+C)) of 48 items, delayed (20 minutes) Free (FR) and Cued (CR) Recall of 16 items, False Recognitions, and Intrusion errors of items.
All subjects gave written informed consent for participation in the study, which was approved by the local ethics committee of the Piti\'e Salp\^etri\'ere Hospital in Paris. All experiments were conducted in accordance with relevant guidelines and regulations.

\begin{table}[h!]
\begin{tabular}{lllllll}
                  & AD (n = 23) &             &    & \multicolumn{3}{l}{HC (n = 25)} \\ \hline
Female/Male       & 12/11       &             &    & \multicolumn{3}{l}{18/7}        \\ \hline
                  & Median      & Range       & NA & Median   & Range         & NA   \\ \hline
Age (years)       & 74          & {[}55,87{]} &    & 69       & {[}34,88{]}   &      \\
Education (years) & 9           & {[}5,22{]}  &    & 14       & {[}5,20{]}    &      \\ \hline
MMSE (/30)        & 23          & {[}15,29{]} &    & 29       & {[}25,30{]}   & 1    \\
TR (F+C)          & 34          & {[}22,47{]} & 2  & 47       & {[}43,48{]}   & 1    \\
CR (delayed) (/16)     & 4           & {[}0,11{]}  & 2  & 12       & {[}7,16{]}    & 1    \\
False Recognition       & 0.5         & {[}0,8{]}   & 1  & 0        & {[}0,1{]}     & 1    \\
Intrusions          & 2           & {[}0,15{]}  & 2  & 0        & {[}0,4{]}     & 1   
\end{tabular}
\caption{Demographic and neurological characterization of subjects. Data is given as Median and Range. Unavailable data are reported as NA. AD: Alzheimer's disease; HC: healthy controls. MMSE: Mini Mental Status Exam; TR (F+C): Total Recall (Free+Cued); CR: Cued Recall.}
\label{tab:data}
\end{table}

The structural brain networks were derived from connectivity matrices based on DWI images, as detailed in \cite{guillon_disrupted_2019}. These matrix elements estimate connection strength (the number of fiber tracts) between $N$ anatomical regions of interest (network nodes) defined by the Lausanne2008 brain atlas~\cite{hagmann_mapping_2008, daducci2012connectome}. In functional brain networks, links represent statistically significant correlations between the BOLD fMRI time series of atlas-defined brain areas. This study focused on wavelet correlation matrices representing functional connectivity in the frequency range 0.05–0.10 Hz~\cite{achard2006resilient}. Notably, both fMRI and DWI networks are referenced to the same atlas, enabling region-specific comparison between the two modalities. Further details on dataset acquisition and preprocessing (DTI and fMRI) are available in \cite{guillon_disrupted_2019}. We constructed networks considering brain atlas parcellations at different scales: $N=82$, $N=128$, and $N=233$ regions of interest (ROIs), corresponding to the number of nodes.

Connectivity matrices are initially associated with fully connected weighted graphs. To optimize the information-to-noise ratio of brain network representations, we applied a threshold proportional filtering, retaining a proportion $0<p<1$ of the links that are stronger in weight. The resulting filtered connectivity matrix is binary. 
As demonstrated in~\cite{de_vico_fallani_topological_2017}, the value of link density that optimizes the trade-off between efficiency (both local and global) and sparsity of brain networks, across a wide variety of conditions and modalities, lies within the range of $[\frac{2}{N-1},\frac{4}{N-1}]$. Accordingly, we selected $p$ to consider networks filtered to a density of $\frac{3}{N-1}$. A more densely connected network would yield a more homogeneous distribution of the score across the nodes, thus reducing localization power. In contrast, for lower densities, several regions would emerge as disrupted. Nevertheless, very low densities may yield an important disruption of the connectivity, beyond the network backbone, which can lead to biased results.

\begin{figure*}[t]
	\includegraphics[width=0.9\linewidth]{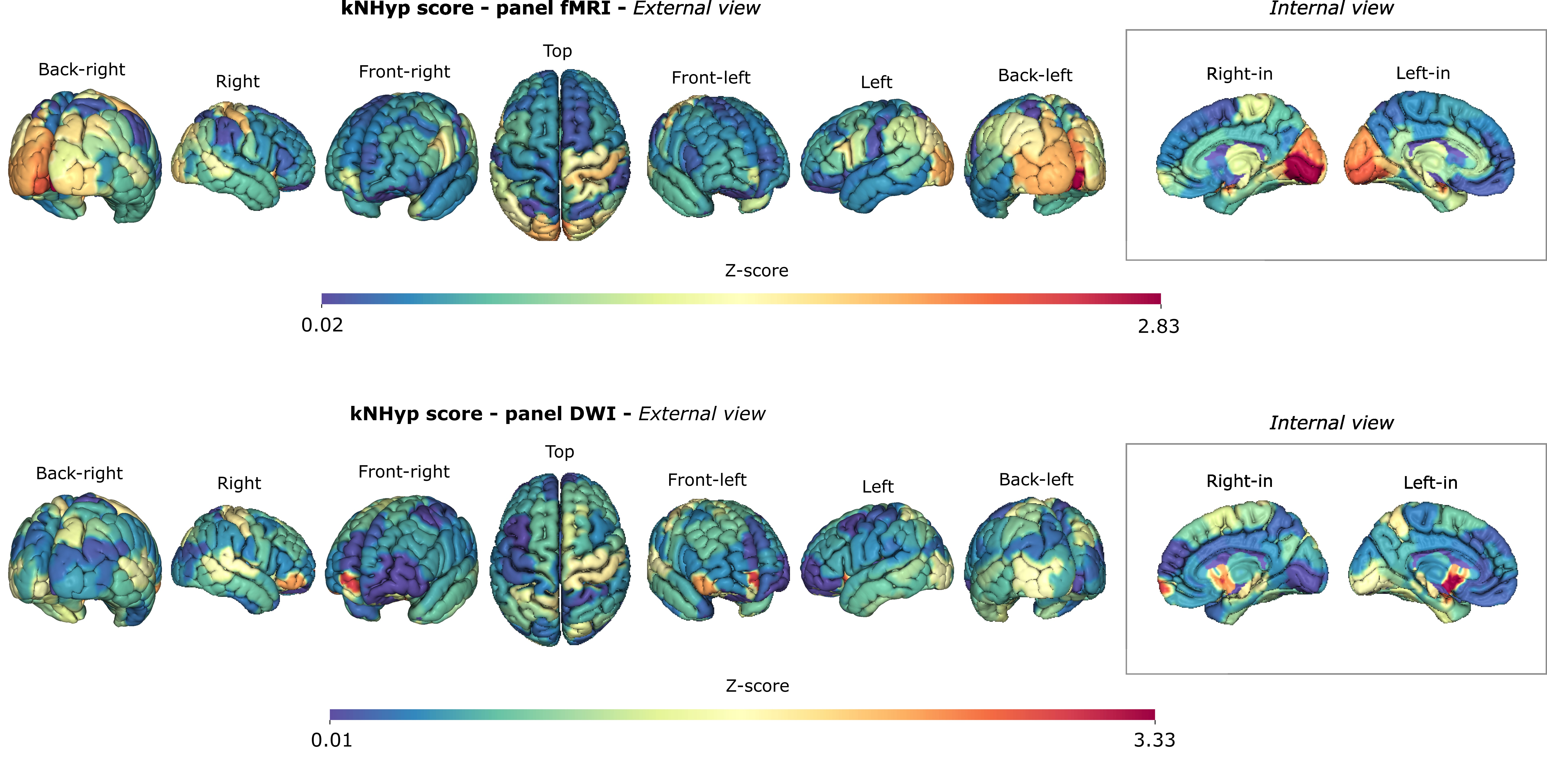}
	\caption{Visualisation of $\mathrm{kNHypD}$ score value on a template brain, obtained from the comparison between the two groups of patients and controls, at a parcellation resolution with 128 regions. Top panel is relative to functional brain networks estimated from fMRI, bottom panel is relative to structural brain networks obtained from DWI imaging.}
	\label{fig:intpanelN128}
\end{figure*}

\subsection{Network embedding in hyperbolic space}
Hyperbolic space is a non-Euclidean space with constant negative curvature. To infer the geometric representations of networks we use a mapping based on the hyperbolic disk, which is a two-dimensional model of hyperbolic geometry. 
The hyperbolic distance between two points in this space, designated by indices $i$ and $j$, with polar coordinates $(r_i,\theta_i)$ and $(r_j,\theta_j)$, can be computed according to the hyperbolic law of cosines:

\begin{multline}   
\label{eq_hyp_distance}
    \cosh \mathrm{d}_{hyp}(i,j) = \cosh r_i \times \cosh r_j \\
    - \sinh r_i \times \sinh r_j \times \cos(\pi-\vert \pi-\vert \theta_i-\theta_j\vert \vert).
\end{multline}

In this study, we employ the coalescent embedding method~\cite{muscoloni_machine_2017} to map brain networks onto the hyperbolic disk. In a nutshell, starting from a binary connectivity graph, the method assigns new effective weights to edges by encoding topological similarities between nodes and using a repulsion-attraction rule that prioritizes edges with a more prominent role in information transmission~\cite{muscoloni_machine_2017}. In particular, we chose the pre-weighting rule:

\begin{equation}
	\omega_{ij} = \frac{d_i+d_j+d_id_j}{1+CN_{ij}},
	\label{kernel}
\end{equation}

where $d_i$ is the degree of node $i$ and $CN_{ij}$ is the number of common neighbors between node $i$ and $j$.

The resulting network $\omega_{ij}$ is then projected onto the two-dimensional hyperbolic disk $D$ by means of nonlinear dimensional reduction techniques, such as Isomap or Laplacian Eigenmaps~\cite{von2007tutorial}. In this study, we applied Laplacian Eigenmaps, as it is computationally efficient and optimally preserves nodes' local neighborhood information~\cite{belkin2003laplacian}. The angular coordinates of the embedded nodes are adjusted uniformly while maintaining their angular order.
Finally, the radius of each node is assigned as a function of its rank when sorted in descending order of node degree: $r_i=\frac{2}{\zeta} (\beta \ln i + (1-\beta) \ln N)$, $i=1,2,...,N$, where $N$ denotes the number of nodes, $\zeta$ is a parameter determining space curvature, and $\beta$ is a fading parameter. Here, we used $\zeta=1$ and $\beta=0.9$. Finally, we rescale the radial coordinates to fit all the embedded networks into the same unitary hyperbolic disk.
Although this transformation is not an isometry, distances are computed just afterward and to estimate the relative changes in neighborhood distances. %

In the context of brain networks, the parameters of similarity and centrality are of significant importance in understanding the role of nodes (electrodes, voxels, or regions of interest) in the structure and function of the system. Nodes can be defined as more "similar" if they have a high number of neighbors in common. It can reasonably be assumed that these nodes are likely to belong to the same anatomical region and have similar functional characteristics. Brain areas with higher centrality typically play a role in integration within the system, interacting with different anatomical or functional regions and acting as centers for information sorting. Within the hyperbolic disk, nodes belonging to the same functional or anatomical module will be angularly close, and nodes with the highest centrality will be closer to the center. The hyperbolic distance in Eq.~\ref{eq_hyp_distance} combines these two dimensions into a unique notion of proximity.

\subsection{\texorpdfstring{$k$}{k}-Neighborhood Hyperbolic Distortion score} \label{subsec:score}
In this study, we use the geometric perturbation score $\mathrm{kNHypD}$, as presented in~\cite{longhena2024detecting}, to compare brain networks based on their embedded configurations. The $k$-Neighborhood Hyperbolic Distortion score ($\mathrm{kNHypD}$) is based on the premise that a local perturbation in the connectivity of a node will modify the similarity relations with its neighbors and, consequently, their relative distances in the embedding space.
To quantify this change at the local (nodal) level, we first define the $k$-neighborhood of node $i$, denoted as $kN(i)$, as the group of the nearest $k$ neighbors of node $i$ in the embedding space of a graph $G$, the hyperbolic disk $\mathbf{D}$. In the embedding space of the perturbed graph $G'$ (denoted here as the hyperbolic disk $\mathbf{D}'$), node $i$ and its neighbors will generally have different coordinates. The $k$-neighborhood of node $i$ in $\mathbf{D}'$ is designated $kN'(i)$, and it generally defines a different set of nodes compared to $kN(i)$. Although the embedding spaces of graphs $G$ and $G'$ have the same metric structure and the same set of embedded nodes, their coordinates differ because the connections they form are different in the original graphs $G$ and $G'$. For each node $i$, we compute the distances between $i$ and the points with index $j \in kN(i)$ in disk $\mathbf{D}$ and in disk $\mathbf{D}'$, respectively. We then compare them using the score function~\cite{longhena2024detecting}:

\begin{equation}
	\mathrm{kNHypD}(i) = \mathlarger{\mathlarger{\sum}}_{j \in kN(i)} \frac{\left|\mathrm{d}_{hyp}(i,j;\mathbf{D}')-\mathrm{d}_{hyp}(i,j;\mathbf{D})\right|}{\mathrm{d}_{hyp}(i,j;\mathbf{D})},
	\label{kNHypD-score}
\end{equation}

\begin{figure*}[!ht]
	\includegraphics[width=0.95\linewidth]{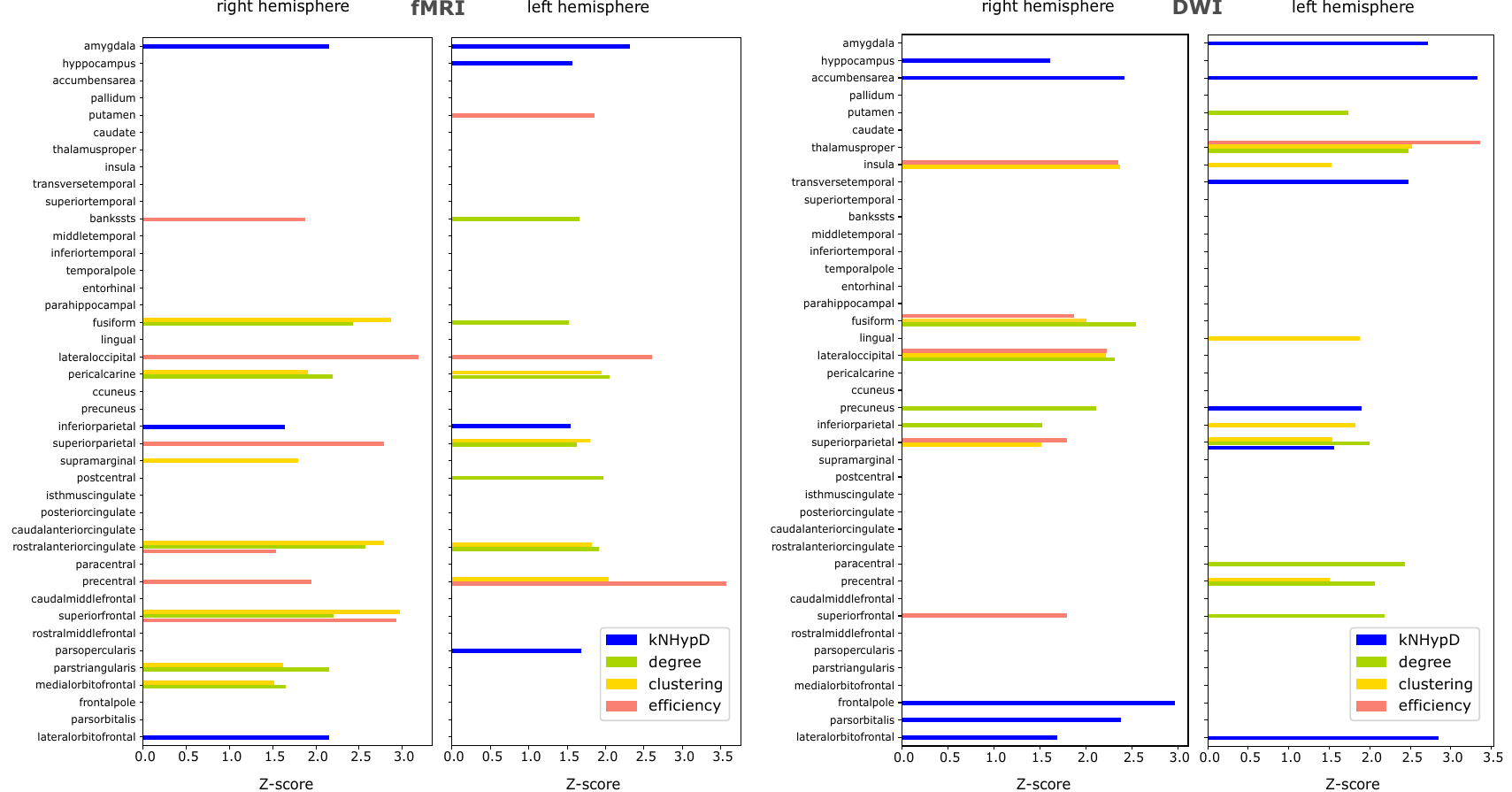}
	\caption{Comparison of brain areas with the most relevant connectivity alterations according to our geometric $\mathrm{kNHypD}$ score and standard topological measures: degree, clustering, and efficiency. The magnitude of the bars represents the number of standard deviations from the average over the whole network. Only values above 1.5 are plotted. The networks considered here have $N=128$ nodes.}
	\label{fig:comparison-kNHypD-topo}
\end{figure*}

Where $\mathrm{d}_{hyp}(i,j;\mathbb{D}')$ denotes the hyperbolic distance (Eq.~\ref{eq_hyp_distance}) between the coordinates corresponding to $i$ and $j$ in the embedding disk $\mathbf{D}'$. The absolute value is introduced to count as a positive contribution both the cases of increase and decrease in the distance to node $i$. The score computation is illustrated in Figure~\ref{fig:method}, using the example of a patient (as a perturbed network) compared with a healthy control.

One of the key advantages of this type of geometric score is its invariance under rotations and reflections of the embedding space. Consequently, if the embedding algorithm introduces a random component to the orientation of the disk, as is the case with the coalescent embedding~\cite{muscoloni_machine_2017}, there is no need to align the embedded networks prior to their comparison.

The best choice of $k$ in the estimation of $\mathrm{kNHypD}$ depends on the data in question. As with other non-parametric estimators in statistics, a small value of $k$ yields a large variance in the score, while a large number of neighbors increases its bias and the computational complexity involved in calculating a large number of terms~\cite{longhena2024detecting}. In accordance with the recommendations set forth in Ref.~\cite{longhena2024detecting}, we consider a neighborhood size of $10\%$ of the total number of nodes $N$. For the studied networks, the results remain consistent when selecting larger neighborhoods.

\section{Results}
\subsection{kNHypD score to detect connectivity disruption in AD} 
We used this geometric measure to characterize connectivity anomalies affecting a group of brain networks in patients diagnosed with Alzheimer's disease compared to a group of healthy controls. In this study, we assume that neurodegeneration has caused a perturbation in the organization of the brain network of each patient compared to the population with healthy characteristics. In this regard, the $\mathrm{kNHypD}$ score can be used to localize and quantify the impact of the perturbation based on the geometric distortion of the embedding.

For each brain network $G_p$ with $p \in \mathcal{P}$ (the group of patients), we compute the corresponding embedding in the hyperbolic disk $\mathbf{D}_p$. The same process is repeated for the group of controls, $\mathcal{C}$, obtaining the set of hyperbolic disk embeddings ${\mathbf{D}_c, c \in \mathcal{C}}$. Then, each patient is compared with each healthy control, following the scheme illustrated in Figure~\ref{fig:method}, resulting in the scores $\mathrm{kNHypD_c^p(i)}$, which estimate the abnormality of each brain region $i$ of patient $p$ with respect to control $c$:

\begin{figure*}[!ht]
	\includegraphics[width=0.95\linewidth]{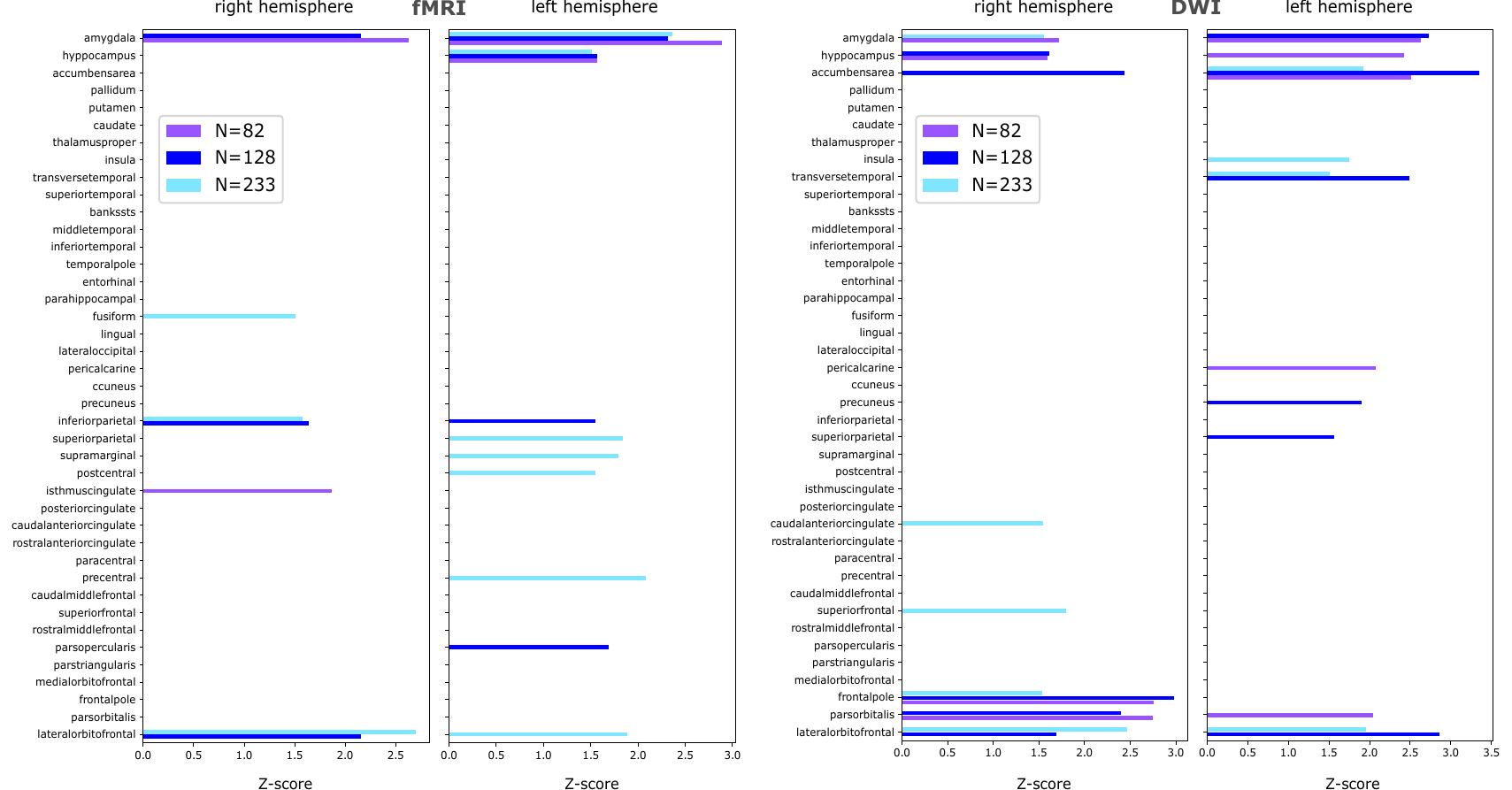}
	\caption{Comparison of anomalous areas identified by the $\mathrm{kNHypD}$ score at three different parcellation scales: 82, 128, and 233 nodes. The results for the latter two scales are coarsened onto the 82 anatomical regions of the lower scale, assigning the maximum value of deviation over the corresponding subregions to each bar. Only bars with magnitudes above 1.5 standard deviations are represented.}
	\label{fig:comparison-scales}
\end{figure*}

\begin{equation}
    \mathrm{kNHypD}^{p}_c(i)= \sum_{j \in kN_c(i)}\frac{\left| \mathrm{d}_{hyp}(i,j;\mathbf{D}_p)-\mathrm{d}_{hyp}(i,j;\mathbf{D}_c) \right| }{\mathrm{d}_{hyp}(i,j;\mathbf{D}_c)}.
\end{equation}

The hyperbolic distance $\mathrm{d}_{hyp}(i,j;\mathbf{D}_p)$ is relative to the coordinates that node $i$ and $j$ assume in the hyperbolic embedding of $G_p$. While $\mathrm{d}_{hyp}(i,j;\mathbf{D}_c)$ is relative to the coordinates they assume in the embedding space of $G_c$. 

We then average over the control group to obtain an absolute set of scores for patient $p$, which represents the patient's local degree of anomaly with respect to the control population:

\begin{equation}
    \mathrm{kNHypD}^{p}(i)= \frac{1}{\left| \mathcal{C}\right|}  \sum_{c \in \mathcal{C}} \mathrm{kNHypD}^{p}_c(i).
\end{equation}

Finally, we averaged across the group of patients to account for intersubject variability and identify the areas of connectivity disruption most indicative of the disease state:

\begin{equation}
    \mathrm{kNHypD}(i)= \frac{1}{\left| \mathcal{P}\right|}  \sum_{p \in \mathcal{P}} \mathrm{kNHypD}^{p}(i).
\end{equation}

It is important to note that each patient's brain network is not being compared with its original, unperturbed configuration prior to the onset of the disease. Instead, it is being compared with a set of control brain networks belonging to different subjects, which will inevitably exhibit differences in the connections, and thus differences in the geometric embedding configurations, due to intersubject variability. This effect introduces a positive non-zero bias in the value of the $\mathrm{kNHypD}$ score estimated at each comparison. To correct for this bias, we identify the regions of the brain exhibiting the highest degree of connectivity perturbation as those exhibiting the highest score relative to the average score calculated over the entire network. The magnitude of the perturbation score will therefore be represented as a Z-score, estimated as the number of standard deviations by which the value deviates from the mean value over the entire network.

The results obtained with the $\mathrm{kNHypD}$ geometric score are compared with those obtained from a t-test between the local (node-specific) values of common connectivity measures (degree, clustering, and efficiency). Figure~\ref{fig:comparison-kNHypD-topo} depicts the most abnormal brain areas identified by the different methods, for both fMRI and DWI modalities, and for networks with a parcellation resolution of $128$ nodes. For visualization purposes, only the areas with Z-score values exceeding 1.5 are shown. The presence of positive Z-scores can only be explained by considering that the most anomalous values are always larger than the mean.

All the metrics considered identify some regions with a relevant abnormality in both anatomical (DWI) and functional (fMRI) networks. These regions belong to the temporal, parietal, and frontal lobes, in accordance with the neuropathology of the disease. We suggest that geometric and topological measures convey complementary information about the disruption in the temporal and parietal lobes. We notice, however, that only the geometric score identifies the amygdala, hippocampus, and the frontal areas (memory and behavioral-related regions) as affected by neurodegeneration.

\subsection{Comparison of different parcellation scales} 
We further compared the anomalous regions identified by the geometric $\mathrm{kNHypD}$ score at three different resolutions of the parcellation. The results are shown in Figure~\ref{fig:comparison-scales}.
Different parcellation scales correspond to different spatial resolutions at which we observe the system and construct its network representation. An anomaly observed at a fine scale could reflect changes in the very short-range connectivity within the subregions of a larger macro-region. Similarly, an anomaly observed at a coarse resolution scale could reflect the sum of the connectivity distortions within all the subregions.

Interestingly, the evidence of connectivity disruption in the amygdala, hippocampus, accumbens area, and the frontal lobe regions is consistent across all the different scales analyzed.
The result for DWI networks is characterized by a more significant distortion in the right frontal lobe, which can be clearly visualized on the brain map in Figure~\ref{fig:intpanelN128}.
Results also show that, in general, different network resolutions contain different information, which is in agreement with the state of the art on brain network analysis across different parcellation scales~\cite{korhonen2017consistency}.

At the lowest resolution of 82 nodes, we uniquely find anomalous values in the isthmus of the cingulate gyrus for the fMRI, and in the pericalcarine region for the DWI structural networks. The DWI networks with 128 nodes additionally show anomalies in the precuneus, superior parietal, and transverse temporal gyrus. In the fMRI networks, on the other hand, the main perturbations are found in the inferior frontal and inferior parietal gyri. As for the largest scale ($N=233$ nodes), we also identify the insula, transverse temporal gyrus, caudal anterior cingulate, and superior frontal cortex in the functional networks, while the fusiform, inferior parietal, superior parietal, supramarginal, postcentral, and paracentral gyri are identified in the anatomical networks.


\section*{Discussion}

The application of network science to neurological problems, enabled by the increasing availability of neuroimaging data acquisition and processing techniques, has proven to be a relevant and non-invasive approach to investigating the state of the brain. Meanwhile, the field of complex networks has seen the introduction of new models and methods for analyzing complex systems. Among these, the study of network geometry~\cite{boguna_network_2021} and methods for embedding graphs in a latent low-dimensional hyperbolic space~\cite{xu2021understanding} provide novel connectivity information inherent in non-Euclidean embedding spaces. Within these frameworks, the metric structure of the original graph, based on topological distance between nodes (e.g., the shortest paths), is replaced by the metric of the latent space, where distances become geodesics in a manifold (spheres, hyperbolic disks, etc.).

In this study, we investigate the advantages of a novel method for studying network perturbations based on geometric embedding of networks in a latent hyperbolic space. We apply it to the task of characterizing connectivity disruption patterns in Alzheimer's disease.
We perform the analysis on two-dimensional representations of brain networks in a latent hyperbolic space, obtained through the coalescent embedding method~\cite{muscoloni_machine_2017}. We then assign a perturbation score, computed in the hyperbolic disk, to each brain region. This method, introduced in~\cite{longhena2024detecting}, is designed to localize perturbed nodes in complex networks and applies to our case study by considering the brain network in disease as a perturbation to the healthy ensemble connectivity.
We have shown that the atrophy process associated with neurodegeneration generates connectivity changes that can be captured in the latent space of brain networks. The proposed score allowed us to identify brain regions with perturbed connectivity in the temporal and parietal lobes, as well as the frontal pole, all of which have been found to be involved in the neurodegeneration process. Notably, the detection of abnormalities in memory-related areas (such as the hippocampus and the amygdala) and in the frontal areas is robust across multiple brain parcellation scales and imaging modalities. Altogether, these results support the hypothesis that AD is a network disease leading to disorganized network configurations at both the anatomical and functional levels~\cite{sanz2010loss, reid2013structural}.

Notably, the medial temporal regions are crucial for memory and spatial navigation. The hippocampus, along with the amygdala, which is also involved in memory formation and emotional processing, is known to be among the first regions to be affected anatomically in Alzheimer's disease. The related clinical symptoms include memory (especially episodic memory) loss and spatial and temporal disorientation. As expected, the patients included in this study exhibited clinically significant signs of memory impairment, as reported in Table \ref{tab:data}. However, the involvement of the frontal lobes, associated with behavioral and social impairments, did not have a clinical counterpart in our sample. Therefore, we can hypothesize that connectivity changes in these regions precede noticeable clinical changes, making our measure a potential biomarker for disease progression.
Clinical biomarkers derived from connectivity are an emerging area of interest. Validation of a new geometric score as a biomarker could be achieved by collecting results from further studies conducted on different datasets, across different imaging platforms, and processing pipelines.
Although theoretical studies support the hypothesis that anatomical brain connections can influence certain aspects of brain dynamics, it is unclear how functional changes might emerge and develop from disconnection patterns initiated by anatomical brain atrophy. Integrating information from structural and functional brain networks in longitudinal studies of healthy subjects progressing to Alzheimer's patients will be fundamental in providing a more comprehensive understanding of neurodegeneration.

Findings suggest that hyperbolic embeddings allow for an efficient representation of brain networks, encoding multiple non-trivial topological information into the node coordinates, particularly combining the notions of node centrality and similarity.
Our score, based on these geometric representations, captures brain connectivity changes more effectively than standard graph topological metrics.
Indeed, we demonstrated that the geometric score conveys complementary information about changes in the parietal and temporal lobes, while uniquely detecting disruptions in memory-related regions and the frontal lobe, which we consider extremely relevant for AD characterization.
We emphasize that the $\mathrm{kNHypD}$ score only involves the computation of relative distances between a certain node and its $k$-neighborhood. In this way, it does not require an alignment of the embeddings to compare, as some other metrics do~\cite{longhena2024detecting} in order to account for their random orientation. This provides an advantage in terms of computational complexity and accuracy.

Moreover, the proposed metric is general enough to capture different types of connectivity changes. Indeed, a non-zero score can result from the focal node or a node in its $k$-neighborhood undergoing: \textit{i)} a degree change, which would cause a radial displacement of the node towards the disk periphery (in the case of edge cut) or towards the center (edge addition); \textit{ii)} an increase or decrease in the angular distance between nodes, reflecting a change in their topological similarity, which can be classified as a rewiring process preserving node degrees; \textit{iii)} a combination of the previous cases.
It's important to note that the $\mathrm{kNHypD}$ score captures the coupled effect of all these types of connectivity changes, since the hyperbolic distance (Eq.~\ref{eq_hyp_distance}) cannot be decomposed into a purely radial and a purely angular component. This gives the score a unique ability to capture complex, multi-factor connectivity perturbations, while implying the impossibility of decoupling contributions from different types of connectivity changes. However, information about how the $k$-neighborhood of a node has changed is encoded in the geometric embeddings. More specifically, we could track the displacement of individual nodes, decompose it into radial and angular components, and compute specific geometric (though not hyperbolic) scores for different types of topological damage. This could allow us to infer whether the damage in a given region is due to disconnection, rewiring, or a combination of these processes. We suggest that future implementations of this approach could include additional features to identify specific types of topological damage, thereby enhancing its utility and interpretability.

We finally note that other non-Euclidean embedding techniques could potentially be combined with the proposed framework to identify local perturbations in the connectivity structure. Alternative hyperbolic mappings (e.g., Mercator~\cite{GarciaMercator1019}, HyperMap~\cite{papadopoulos2014network}, or Hydra~\cite{keller2020hydra}, among others) might be worth considering for the representation of brain networks in a lower-dimensional space. Similarly, other dimensionality reduction methods (e.g., Isomap, eigenmaps, etc.) could be integrated into the coalescent embedding method to capture the neighborhood information of a network's nodes~\cite{von2007tutorial}. The choice of the method can be made depending on the task and demands of the analysis to be performed.

In conclusion, our method effectively captures brain disruptions at the network level, identifying areas likely responsible for deficits affecting the patient at the time of the study, as well as regions that may already be affected but clinically silent. We demonstrate that simple measures in the hyperbolic embedding space are able to capture the typical areas affected by neurodegeneration. Latent hyperbolic space representations of brain networks promise to be the starting point of analyses that could improve our understanding of the brain in health and disease. A longitudinal study could further verify the hypothesis that the functional alteration of the frontal areas detected here precedes atrophy and the onset of clinical behavioral signs in patients.
The proposed method is flexible enough to be applied to other clinical cases, namely epilepsy, schizophrenia, Parkinson’s disease, and many others. Future studies could help assess the power of hyperbolic latent space representations of networks in localizing subtle yet complex connectivity perturbations, which could be crucial in the characterization and diagnosis of different neurological diseases.

\section{Acknowledgments}
We have benefited immensely from discussions with many colleagues and friends in the last few months, but we would like to thank especially Vincent Le Du for his help in the visualization of the results with Visbrain~\cite{visbrain-repo,visbrain-NERV-repo}. A.L. acknowledges financial support from the doctoral school EDITE at Sorbonne University, Paris (FR).



%

\end{document}